\documentclass[12pt]{article}
\usepackage{amsmath,amsfonts}
\usepackage{graphicx}
\usepackage{natbib}
\usepackage{url} 
\usepackage{tikz}
\usepackage{multirow}

\newcommand{\blind}{0}

\begin{document}

\def\spacingset#1{\renewcommand{\baselinestretch}%
{#1}\small\normalsize} \spacingset{1}


\if0\blind
{
  \title{\bf Ordinal Probit Functional Outcome Regression with Application to Computer-Use Behavior in Rhesus Monkeys}
  \author{Mark J. Meyer\thanks{
    This work was supported by grants from the National Institutes of Health (ES-007142, ES-000002, CA-134294, CA-178744, R01MH081862), the National Science Foundation (\#IOS-1146316), and ORIP/OD (P51OD011132).}\hspace{.2cm}\\
    Department of Mathematics and Statistics, \\Georgetown University\\
    and \\
    Jeffrey S. Morris \\
    Department of Biostatistics, Epidemiology, and Informatics, \\Perelman School of Medicine, University of Pennsylvania\\
    and\\
    Regina Paxton Gazes\\
    Department of Psychology and Program in Animal Behavior, \\Bucknell University\\
    and\\
    Brent A. Coull\\
    Department of Biostatistics, \\Harvard T. H. Chan School of Public Health}
  \maketitle
} \fi

\if1\blind
{
  \bigskip
  \bigskip
  \bigskip
  \begin{center}
    {\LARGE\bf Historical Functional Linear Models with Wavelet-packets}
\end{center}
  \medskip
} \fi

\bigskip
\begin{abstract}
Research in functional regression has made great strides in expanding to non-Gaussian functional outcomes, but exploration of ordinal functional outcomes remains limited. Motivated by a study of computer-use behavior in rhesus macaques (\emph{Macaca mulatta}), we introduce the Ordinal Probit Functional Outcome Regression model (OPFOR). OPFOR models can be fit using one of several basis functions including penalized B-splines, wavelets, and O'Sullivan splines---the last of which typically performs best. Simulation using a variety of underlying covariance patterns shows that the model performs reasonably well in estimation under multiple basis functions with near nominal coverage for joint credible intervals. Finally, in application, we use Bayesian model selection criteria adapted to functional outcome regression to best characterize the relation between several demographic factors of interest and the monkeys' computer use over the course of a year. In comparison with a standard ordinal longitudinal analysis, OPFOR outperforms a cumulative-link mixed-effects model in simulation and provides additional and more nuanced information on the nature of the monkeys' computer-use behavior.
\end{abstract}

\noindent%
\vfill

\newpage
\spacingset{1.5} 

\section{Introduction}
\label{s:intro}
\vspace{-10pt}
\cite{Gazes2013} present a study of computer-use patterns in a socially-housed group of rhesus macaques (\emph{Macaca mulatta}) at the Yerkes National Primate Research Center in Atlanta, GA who were given access to automated touch-screen computer systems between March 2009 and April 2014. Computer-use data was collected for all non-infant monkeys in the colony. Each animal had an RFID implant that was read when the monkey used the computer, and used to track individual computer use. The raw data is the daily amount of time each monkey spent using the touch-screens to complete tasks ranging from size and shade discrimination to matching images. \cite{Gazes2013} give specific details on the tasks. Although the raw data can be considered continuous, the broader trends of computer use on an ordinal scale, such as no use, low use, moderate use, and high use, have more practical value for psychologists in choosing productive research subjects. These classifications are particularly important for assessing what types of comparative research questions can be asked in the population. The resulting outcome is an ordinal variate that varies daily when measured over the course of a year. Standard longitudinal approaches could be used on this data but would not fully capture the seasonal nature of computer-use. Instead, a functional outcome regression framework is needed that can accommodate an ordered outcome.

In this paper, we introduce the Ordinal Probit Functional Outcome Regression (OPFOR) model, a fully Bayesian approach to regressing an ordinal functional outcome on a matrix of scalar covariates. The modeling framework uses the data-augmented probit formulation of ordinal data to represent an ordinal functional outcome as a latent Gaussian functional outcome. This representation allows us to consider and compare multiple choices of basis functions including wavelet-bases, B-spline bases, and O'Sullivan splines (O-splines). Joint credible intervals have previously been used for inference in Bayesian functional regression models, and we examine these intervals in our ordinal functional framework. Estimation uses one of two Markov chain Monte Carlo algorithms, depending on the choice of basis; we provide MATLAB code, available at \url{https://github.com/markjmeyer/OPFOR}.

Our work is motivated by the computer-use data which we analyze using OPFOR to determine the nature of the relation between demographic factors and computer use over the course of a year. To fully identify this relation, we implement Bayesian model-selection tools adapted to the functional case including the deviance information criterion (DIC), the log posterior predictive density (lppd), and the Watanabe-Akaike information criterion (WAIC) \citep{Gelman2013}. We use these criteria to select the best model for the computer-use data and conduct posterior functional inference using joint intervals. The application demonstrates the modeling frameworks ability to handle ordinal functional data sampled on a relatively large grid with a general matrix of scalar covariates. Finally, a comparison analysis using cumulative-link mixed-effects models (CLMM), a standard ordinal longitudinal regression approach, suggests that the OPFOR analysis better captures the relation and provides additional insights a standard analysis would have missed. In addition, we show that, in simulation, our approach outperforms the CLMM in capturing a range of effects that mimic the nature of the computer-use patterns in the monkeys over the course of the year.

Section \ref{s:data} contains more detail on the scientific motivation as well as the statistical motivation and literature review. Section \ref{s:opfm} presents the general modeling framework for OPFOR. Section \ref{s:latent} gives details on both the spline- and wavelet-based models. In Section \ref{s:sim}, we discuss the results of our simulation study. Section \ref{s:app} describes our findings from the analysis of the computer-use data using OPFOR. And in Section \ref{s:disc}, we provide a discussion of the methodology and the data analysis.

\section{Scientific and Statistical Motivation}
\label{s:data}

Previous longitudinal analyses of touch-screen computer-use in rhesus macaques reveal important demographic patterns that are practically useful for identifying which subjects produced adequate data, and scientifically interesting for understanding social behavior in monkeys \citep{Gazes2019}. For example, adult female monkeys decrease their engagement with the touch-screen systems during the breeding and birthing seasons, and longitudinally after giving birth to their first child \citep{Gazes2019}. This suggests that female monkeys will forgo touch-screen computer use in light of new social responsibilities such as breeding and childcare. Although the results of these analyses indicated some effects of seasonality as well as demographic factors---including age, sex, and social dominance rank---on touch-screen use, standard longitudinal techniques could not fully capture the nature of these associations. The present study uses the data set to identify factors that are associated with how much a monkey uses the system on a given day---categorized as no, low, moderate, and high use---and how that use changes over the course of the year (breeding season, winter, birthing season, summer). Classifying computer use on an ordinal scale is useful for studies of cognitive performance based on dominance rank in a population. But if low ranking animals show low or no use during particular seasons, the resulting data may be insufficient to relate performance to dominance rank. Thus, identifying demographic features that impact seasonal computer-use would be beneficial.

Daily use of the touch-screens is an ordinal variate measured serially over time. For such an ordinal functional outcome we develop OPFOR, a function-on-scalar regression model, in which the outcome is a function of time and the predictors are scalars, to model how demographic-specific computer use varies over the course of a year. This will allow us to better identify demographic groups that produce sufficient subject numbers for cognitive testing, and inform understanding of the behavior of socially living monkeys generally. Our method is also timely and important for the field of psychology, as it provides a framework for functional analysis of data from automated cognitive testing systems for animals housed in large social groups, a relatively new type of testing that is increasing in popularity through advances in technology \citep{Fagot2010,Gazes2019}. In the statistical literature, this model has received limited attention despite the volume of work on functional outcomes.

\cite{Faraway1997} and \cite{RamsaySilverman1997} represent the early work on function-on-scalar regression, but the literature has grown. For example, \cite{Guo2002}, \cite{ShiEtAl2007}, and \cite{ReissEtAl2010} consider kernel smoothing and spline-based approaches to model a functional outcome, and \cite{KraftyEtAl2008} and \cite{Scheipl2015} explore the use of functional principal components. \cite{Morris2015} provides an extensive review of the function-on-scalar regression literature. In the Bayesian context, \cite{MorrisCarroll2006} introduce Wavelet-based Functional Mixed Models (WFMM) for function-on-scalar regression, and \cite{Goldsmith2016} use splines and functional principal components (fPC) in both fully Bayesian and variational Bayesian models. Both approaches provide flexible frameworks for modeling functional outcomes in a number of settings, and several authors extend the two methodologies. \cite{Zhu2011,Zhu2012} discuss robust adaptive regression and classification in WFMMs, respectively, and \cite{Meyer2015} introduce the function-on-function extension. \cite{Lee2018} extend the WFMM to semi-parametric models with smooth nonparametric covariate effects in function-on-scalar regression models, with smoothing by O'Sullivan splines \citep{Wand2008}. Extensions of \cite{Goldsmith2016} include \cite{Chen2016} and \cite{Goldsmith2017}, which examine variable selection in function-on-scalar and functional linear concurrent models, respectively. However, both frameworks can only be applied to continuous-valued functional responses.

Although the literature on generalized function-on-scalar regression has expanded, it remains limited in scope. \cite{Hall2008} extend functional principal components analysis to accommodate generalized outcomes.  \cite{Li2014}, \cite{Gertheiss2015}, \cite{Scheipl2016}, and \cite{Li2018} develop frequentist methods for binary and count functional outcomes using GEE-type, maximum-likelihood-based, quasi-likelihood-based, and fPC-based approaches, respectively. \cite{vanderLinde2009} proposes a Bayesian version of the work of \cite{Hall2008} using a variational algorithm to conduct approximate Bayesian inference for count and binary data. \cite{vanderLinde2011} extends this approach further, proposing reduced-rank models for functional canonical correlation analysis and functional discriminant analysis. \cite{Wang2014} implement an empirical Bayesian learning approach using a Gaussian approximation and B-splines for outcomes arising from exponential families. \cite{Goldsmith2015} develop a fully Bayesian model for multilevel function-on-scalar regression using fPC and penalized splines for binary and count data.

Much of the existing literature focuses on either the binary or the Poisson case. The method of \cite{Wang2014} is capable of handling an ordinal functional outcome. However, it is not the primary focus of their work, and the authors only discuss a single simulated data setting without exploring the model's operating characteristics in that context. Also no publicly available code implements their method. To our knowledge, no previous work implements probit-based modeling in functional outcome regression. Moreover, wavelet bases and O'Sullivan splines have yet to be examined in generalized functional outcome regression models, let alone in the ordinal case---O'Sullivan splines in particular have received limited attention in the functional literature, although \cite{Wand2008} show that O'Sullivan splines can be more efficient than B-splines.

\section{Ordinal Probit Functional Outcome Regression Model}
\label{s:opfm}

Let $Y_i(t)$ be the observed value for subject $i$ at time $t$, $i = 1,\ldots,N$ and $t = t_1,\ldots, t_T$ where $Y_i(t)$ takes on one of $L$ possible values: $\ell = 0, 1, \ldots, L-1$. In the context of the computer-use data, the values of $Y_i(t)$ are 0 (none), 1 (low), 2 (moderate), and 3 (high) depending on a monkey's level of computer-use on days $t_1 = 1, t_2 = 2, \ldots, t_{365} = 365$. As $t$ only indexes measurement time, $Y_i(t)$ need not necessarily be sampled on an equally spaced grid; however, we do assume that each subject has the same number of measurements taken at approximately the same intervals. Without loss of generality, let $x_i$ represent a single scalar covariate for subject $i$. Extension to a vector of scalar covariates is straightforward. 

We represent $Y_i(t)$ by a continuous latent process, $Y_i^*(t)$. The unobservable $Y_i^*(t)$ can be considered additional data. Thus, this approach is referred to as a data augmentation strategy \citep{AlbertChib1993}. In this work we extend this strategy to functional outcomes. The behavior of $Y_i(t)$ is specified by the relationship
The behavior of $Y_i(t)$ is specified by the relationship
\begin{align}
	Y_i(t) = \left \{ \begin{array}{cl}
		0 & \text{if } c_0 < Y_i^*(t) < c_1\\
		1 & \text{if }c_1 \leq Y_i^*(t) < c_2\\
		 & \vdots \\
		\ell & \text{if }c_{\ell} \leq Y_i^*(t) < c_{\ell+1}\\
		 & \vdots \\
		L-1 & \text{if } c_{L-1} \leq Y_i^*(t) < c_L,
		\end{array} \right.
	\label{m:latentMap}
\end{align}
for $c_0 = -\infty$, $ c_L = \infty$, and the cut points $c_1, \ldots, c_{\ell}, \ldots, c_{L-1}$, satisfying $-\infty < c_1 < \ldots < c_{\ell} < \ldots < c_{L-1} < \infty$. The probability that $Y_i(t) = \ell$ is
\begin{align}
	P\{Y_i(t) = \ell\} = P\{Y_i^*(t) \in [c_{\ell}, c_{\ell+1}) \},\ \ell = 1, \ldots, L-1,
	\label{m:genProb}
\end{align}
with $P\{Y_i(t) = 0\} = P\{Y_i^*(t) \in (c_{0}, c_{1}) \}$. $Y^*(t)$ can be modeled as a functional outcome arising from a continuous distribution.

Let the form of $Y_i^*(t)$ be the functional outcome regression model
\begin{align}
	Y_i^*(t) &= x_i \beta(t) + E_i(t).
	\label{m:latent}
\end{align}
The predictor, $x_i$, is a time-invariant covariate, while $\beta(t)$ is the population averaged effect of $x_i$ on $Y_i(t)$. In the context of the computer-use data, $x_i$ represents a demographic factor of interest and $\beta(t)$ is the corresponding population averaged effect on computer-use measured at time $t$. Assuming Gaussian process errors and standardizing $Y_i^*(t)$, we can express $P\{Y_i(t) = \ell\} $ in terms of the CDF of the standard Gaussian, $\Phi(\cdot)$. The probability at time $t$ for subject $i$ is then
\begin{align}
	P[Y_i(t) = \ell] = \Phi\left[c_{\ell+1} -x_i \beta(t)\right] - \Phi\left[c_{\ell} - x_i \beta(t)\right]
	\label{m:pprobs}
\end{align}
for $\ell = 0, 1, \ldots, L-1$. This assumption results in a probit model.

Models \eqref{m:latent} and \eqref{m:pprobs} represent the formulation for continuous functions; however, we only observe discrete realizations of $Y^*_i$. Assuming all subjects have the same number of measurements and no missing values, we stack the $Y^*_i$ into an $N\times T$ matrix ${Y}^*$. The discrete version of model \eqref{m:latent}, across all subjects, is then
\begin{align}
	{Y}^* = X\beta + {E},\ {E} \sim \mathcal{MN}(0,I_{N\times N},{\it\Sigma}_{{E}}),
	\label{m:discLatent}
\end{align}
where $\mathcal{MN}$ denotes the Matrix Normal distribution \citep{DeWaal2014}, $I_{N\times N}$ is an $N\times N$ identity matrix representing the covariance of the rows, ${\it\Sigma}_{{E}}$ is the $T\times T$ covariance matrix of the columns, and ${E}$ is $N\times T$. If $P$ is the number of covariates, then ${\beta}$ is a $P\times T$ matrix and $X$ is an $N \times P$ matrix. Regardless of the basis expansion used, $Y^*$ and, if desired, the $c_{\ell}$ are sampled in the same fashion. After fixing the first cut point, $c_1$, typically at 0, we assume a flat prior on the remaining $c_{\ell}$, $\ell = 2, \ldots, L-1$. For identifiability, the first cut point must be fixed. The remaining cut points may be sampled or fixed, depending on the data construction. The covariance of the columns of $Y^*$, ${\it\Sigma}_{{E}}$, is also not fully identifiable because the latent variable representation is invariant to scalar transformations. As in the univariate case, we set the variances to one, letting ${\it\Sigma}_{{E}} = I_{T\times T}$. Nevertheless, we still sample ${\it\Sigma}_{{E}}$, with estimation depending on basis choice. We do not retain these samples, however, and only allow them to inform our estimation of the model coefficients. Conditional posterior distributions for $Y^*$ and the $c_{\ell}$ are described in the Supplementary Material.

\section{Modeling the Latent Functional Outcome}
\label{s:latent}

Given the data augmented in \eqref{m:discLatent}, we can model ${Y}^*$  as a matrix of Gaussian functional outcomes. Typically, the functional form of $\beta$ is modeled using a basis expansion. Two commonly used types of basis functions are wavelets, as in \cite{MorrisCarroll2006}, and B-splines, which \cite{Goldsmith2016} use. We develop two approaches in the spirit of this prior work, utilizing both types of basis functions and ultimately propose the use of O-splines for OPFOR. We also formulate two existing interval estimation techniques in the context of our model and describe a model selection criterion.

\subsection{Wavelet-based Model Formulation}
\label{ss:wave}

Using model \eqref{m:discLatent} and applying a Discrete Wavelet Transformation (DWT) to $Y^*$ gives the decomposition ${Y}^*= {Y}^{*W}{W}$ where ${Y}^{*W}$ contains the wavelet-space coefficients and ${W}$ is a $T^*\times T$ matrix of wavelet basis functions for $T^*$ coefficients. For additional details on the DWT, see the Supplementary Material. Decomposing ${\beta}$ gives ${\beta} = {\beta}^W  {W}$ and the DWT applied to ${E}$ is ${E} = {E}^W {W}$. The model is then
\begin{align}
	{Y}^{*W}{W} = X{\beta}^W  {W}+ {E}^W {W}.
	\label{m:latWavExp}
\end{align}
The rows of the matrix of the wavelet basis functions are orthogonal, thus ${W} {W}' = {I}_{T^*}$. Post-multiplying model \eqref{m:latWavExp} by ${W}'$ gives
\begin{align}
	{Y}^{*W} = X {\beta}^W + {E}^W,
	\label{m:wavLatent}
\end{align}
where ${E}^W \sim \mathcal{MN}(0, I_{T^* \times T^*}, {\it\Sigma}_{{E}^W})$ for ${\it\Sigma}_{{E}} = W'{\it\Sigma}_{{E}^W}{W}$.

	Wavelet coefficients are double-indexed by their scale, $j$, and location, $k$. For an $N\times P$ matrix of covariates, $X$, assuming independence in the wavelet space enables the sequential fitting of separate models of the form
\begin{align}
	\textbf{y}_{(j,k)}^{*W} = X\boldsymbol{\beta}_{(j,k)}^{W} + \textbf{e}_{(j,k)}^{W},
	\label{m:priorLatent}
\end{align}
where $\textbf{y}_{(j,k)}^{*W}$ and $\textbf{e}_{(j,k)}^{W}$ are $N\times1$ vector components of ${Y}^{*W}$ and ${E}^W$, respectively, and $\boldsymbol{\beta}_{(j,k)}^{W}$ is a $P\times1$ vector of coefficients. We place spike and slab priors on the coefficients $\boldsymbol{\beta}_{(j,k)}^{W} = \left\{ \beta_{(p, jk)}^{W} \right\}$, where $p$ indexes the columns of $X$. The prior on the $p$th coefficient from model \eqref{m:priorLatent} is
\begin{align}
	 \beta_{(p, jk)}^{W} \sim \gamma_{(p,jk)}\mathcal{N}(0, \tau_{pj}) + (1-\gamma_{p,jk})d_0,\ \ \gamma_{(p,jk)} \sim Bern(\pi_{pj}),
\end{align}
where $d_0$ is a point-mass distribution at zero. This prior performs adaptive shrinkage in the wavelet space. The independence assumption and spike and slab priors are consistent with the existing literature on wavelet-based functional regression \citep{MorrisCarroll2006,MalloyEtAl2010,Zhu2011,Meyer2015}. On $\tau_{pj}$ and $\pi_{pj}$, we place the hyper-priors  $\tau_{pj} \sim {IG}(a_{\tau},b_{\tau})\text{ and }\pi_{pj} \sim Beta(a_{\pi},b_{\pi})$, where the hyper-parameters $a_{\tau},b_{\tau}, a_{\pi},$ and $b_{\pi}$ are fixed and based on empirical Bayes estimates. Finally, we place a mean zero normal prior on each $\textbf{e}_{(j,k)}^{W}$ with variance $\sigma^2_{(j,k)}$. The covariance ${\it\Sigma}_E$ encompasses a broad class of covariance structures, as shown in \cite{MorrisCarroll2006}. Full conditionals for the model under these prior specifications are provided in the Supplementary Material.

\subsection{Spline-based Model Formulation}
\label{ss:spline}

Let $\Theta$ denote a $T\times K$ matrix of basis functions and $\beta^S$ be a $K\times P$ matrix of basis coefficients such that $\beta = \left(\Theta\beta^S\right)'$. To model between curve covariance, we use fPC with $K_E$ scores. Let $E = C(\Theta\beta^{E})' + E^S$ where $C$ is an $N \times K_E$ matrix of subject scores, $\beta^E$ is the $K\times K_E$ matrix of coefficients, and $E^S$ is an $N\times T$ matrix of independent errors. The spline-based model is then
\begin{align}
	{Y}^* = X{\beta^S}'\Theta' + C{\beta^E}'\Theta' + E^S,
	\label{m:spline}
\end{align}
for the $N\times P$ matrix of scalar covariates, $X$.

We penalize the fits of both $\beta^S$ and $\beta^E$ using a penalty matrix, $\Delta$, that is dependent upon choice of spline basis. For B-spline-based models, we use the penalty matrix \cite{Goldsmith2016} use: $\Delta = \xi\Delta_0 + (1-\xi)\Delta_2$ where $\Delta_0$ and $\Delta_2$ are the zeroth and second derivate penalties, respectively. The control parameter $\xi$ is between 0 and 1 and chosen to strike a balance between smoothness and shrinkage; values near zero favor shrinkage. We use $\xi = 0.01$ for all B-spline-based models.

\cite{Wand2008} show that O-splines can be more efficient than B-splines, requiring fewer basis functions for similar fits. To construct O-splines, we begin with the standard B-spline expansion using $K = O + 4$ knots. The $(d, d')$ entry of the penalty matrix is $\int_a^b \ddot{\theta}_d(x)\ddot{\theta}_{d'}(x)dx$ where $\ddot{\theta}_d$ is the second derivative function of the $d$th basis function for $x \in \mathbb{R}$. The bounds, $a$ and $b$, are the end points of a non-decreasing sequence of knots. Using O-splines in a mixed model framework corresponds to penalizing the fit with $\lambda_d \int_a^b {\left\{\ddot{\theta}_d(x)\right\}}^2dx$, where $\lambda_d$ is the tuning parameter for the $d$th element of the diagonal of $\Delta$.

Regardless of spline type, we place mean zero normal priors on $\beta^S$ and $\beta^E$ with covariances equal to $(\Lambda_S \otimes \Delta)^{-1}$ and $(\Lambda_E \otimes \Delta)^{-1}$, respectively, where $\otimes$ denotes the Kronecker product. Both $\Lambda_S$ and $\Lambda_E$ are diagonal matrices where the diagonal elements are equal to $\lambda_{p, S}^{-1}$ and $\lambda_{k_E, E}^{-1}$ for the $p$th column of $\beta^S$ and the $k_E$th fPC score, for the scores $1, \ldots, k_E, \ldots K_E$. We also place a mean zero normal prior on each row of $C$ with variance $I_{K_E}$. Finally, the rows of $E^S$ are independent and normally distributed with mean zero and variance $\sigma_E^2I_{T}$. These specifications are consistent with prior work on spline-based function-on-scalar regression \citep{Goldsmith2015,Goldsmith2016}. The full conditionals for the spline-based model are provided in the Supplementary Material.

\subsection{Interval Estimation and Model Selection}
\label{ss:inf}
We construct the joint-credible intervals \cite{Meyer2015} and \cite{Lee2018} implement, which are similar to those described in \cite{Ruppert2003}. Let $\widehat{\beta}(t)$ and $\widehat{\text{St.Dev}}\left\{ \widehat{\beta}(t) \right\}$ denote the mean and standard deviations of the posterior samples of $\beta(t)$. The interval is $\widehat{\beta}(t) \pm q_{(1-\alpha)}\left[ \widehat{\text{St.Dev}}\left\{ \widehat{\beta}(t) \right\} \right]$, where $q_{(1-\alpha)}$ is the $(1-\alpha)$ quantile taken over the $B$ MCMC samples of the quantity $Z^{(b)}=\ \max_{t \in \mathcal{T}}\left| \frac{\beta^{(b)}(t) - \widehat{\beta}(t)}{\widehat{\text{St.Dev}}\left\{ \widehat{\beta}(t) \right\}} \right|$ for $b = 1, \ldots, B$. 

To perform model selection, we adapt three Bayesian selection criteria to functional outcome regression.
We calculate the adapted lppd using
\begin{align*}
	lppd = \sum_{i=1}^N \log{\left[ \frac{1}{B} \sum_{b=1}^B \prod_{t = t_1}^{t_T} \mathcal{L}\left\{Y_i(t) | \psi^{(b)} \right\} \right]},
\end{align*}
where $\mathcal{L}\{\cdot\}$ denotes the likelihood and $\psi^{(b)}$ represents the $b$th posterior sample of the vector of model parameters. WAIC is a penalized version of lppd and has the form $WAIC = -2 lppd + 2p_{W}$. The penalty term, $p_{W}$, equals
\begin{align*}
	p_{W} = \sum_{i=1}^N \frac{1}{B-1}\sum_{b=1}^B \left[ \sum_{t = t_1}^{t_T} \ell\left\{Y_i(t) | \psi^{(b)}\right\} - \frac{1}{B}\sum_{j=1}^B \sum_{t = t_1}^{t_T} \ell\left\{Y_i(t) | \psi^{(j)} \right\} \right]^2,
\end{align*}
where $\ell\{\cdot\}$ is the log likelihood. The formula for the adapted DIC is in the Supplementary Material. As with other information criteria such as AIC and BIC, the model with the lowest WAIC is considered the best model. WAIC has an advantage over DIC in that it uses the entire posterior sample whereas DIC relies on a point estimate \citep{Gelman2013}. WAIC also asymptotically approximates leave-one-out cross validation. Thus, we prefer it for selection in our modeling context.

\section{Simulation Study}
\label{s:sim}

To investigate the operating characteristics of our method, we develop four ``true'' curve settings to simulate ordinal functional outcomes from a single covariate, $x_i$, which we draw from a standard normal distribution. The four curves are referred to as the sigmoidal, seasonal, decay, and peak settings. These settings suggest different patterns of computer use in rhesus macaques over the course of the year. For example, in the sigmoidal setting, changes in use associated with a positive change in $x_i$ slowly increase over the course of the year while in the peak setting, computer use crests midway through the year and attenuates toward zero after. For a given setting, we generated curves $N = 40$ subjects consisting of $t_{365} = 365$ ordinal variates using one of three underlying covariance structures: independent (least realistic), exponential (most realistic), and compound symmetric, also called exchangeable. The curves consisted of randomly generated outcomes from a multinomial distribution with $L = 4$ categories. Functions describing each curve setting and the covariance structures are in the Supplementary Material.

Combining the four settings with the three covariance structures gives us a total of 12 simulated scenarios. For each, we generate 200 datasets and run six models using B-splines with $K = 5$ and 10 knots (as in \cite{Goldsmith2016}), O-splines with $O = 2$ and 4 knots, and Symmlets with 8 vanishing moments and $J = 6$ and 8 levels. We take 1000 total posterior samples for each model, discarding the first 500. Runtimes for the full 1000 samples vary by basis type but the spline-based models typically took between 12 and 14 seconds. Wavelet models took considerably longer running between 94 and 107 seconds. OPFOR models were run on a laptop with a 2.9 GHz Intel Core i5 processor with 16 GB of memory. The computation was performed using MATLAB version R2017a.

Because the computer-use data has perviously been analyzed using more standard longitudinal regression models, we compare the OPFOR to a standard ordinal longitudinal approach in simulation. Thus for each scenario, we also fit a cumulative-link mixed-effects model (CLMM) with a probit link \citep{Agresti2013}. The CLMM has as its fixed effects the simulated $x_i$ and time, and uses subject-specific random intercepts and random slopes of time to account for within-subject variability. Unlike the OPFOR, the CLMM does not produce estimates of functional effects. However, it can estimate the mean of the outcome. Let $\eta_i(t)$ be the mean of the outcome from~\eqref{m:latent}, $\eta_i(t) = x_i\beta(t)$. We then estimate $\eta_i(t)$ using the six OPFOR models and the CLMM. The longitudinal models are fit using the \texttt{clmm} function in the \texttt{ordinal} package in \texttt{R} \citep{Ordinal2019} on the same laptop with run times varying between 25 and 35 seconds.

Due to space constraints, we only present results for the sigmoidal and peak settings. OPFOR models produce functional estimates and can be compared using $\beta$-\emph{MISE} $= \int [\beta(t) - \hat{\beta}(t)]^2 dt$. To compare OPFOR to the CLMM, we find the average MISE for the mean of the outcome using $\eta$-\emph{MISE} $= \frac{1}{N} \sum_{i=1}^N \int [\eta_i(t) - \hat{\eta}_i(t)]^2 dt$. Table~\ref{t:beta} contains $\beta$-\emph{MISE} values averaged over all 200 simulated datasets while Table~\ref{t:eta} contains the same for $\eta$-\emph{MISE}.

\begin{table}
\centering
\caption{\it $\beta$-MISE for each choice of OPFOR basis function under the sigmoidal and peak scenarios averaged over all 200 datasets. Bold-faced entries denote the lowest value at each Setting and Covariance Structure, Cov. Str., combination. Italicized entries denote the second smallest value. The abbreviations Ind., Exp. and, C. S. refer to Independence, Exponential, and Compound Symmetric, respectively.}
\label{t:beta}
\begin{tabular}{llcccccc}
  \hline
  \hline
 & \it Cov. & \multicolumn{2}{c}{\it B-Spline} & \multicolumn{2}{c}{\it O-Spline} & \multicolumn{2}{c}{\it Symmlets} \\
\it Setting & \it Str. & $K = 5$ & $K = 10$ & $O = 2$ & $O = 4$ & $J = 6$ & $J = 8$ \\ 
  \hline
  \hline
Sigmoidal & Ind. & 0.0022 & 0.0014 & \it0.0010 & 0.0012 & \bf 0.0009 & 0.0033 \\ 
 & Exp. & 0.0024 & 0.0017 & \bf 0.0012 & \it0.0014 & \bf 0.0012 & 0.0035 \\ 
 & C. S. & 0.0023 & 0.0016 & \bf 0.0010 & 0.0013 & \it0.0011 & 0.0034 \\ 
   \hline
Peak & Ind. & 0.0110 & \it0.0014 & 0.0021 & \bf0.0011 & \bf0.0011 & 0.0024 \\ 
  & Exp. & 0.0111 & \it0.0015 & 0.0021 & \bf0.0012 & \bf0.0012 & 0.0025 \\
  & C. S. & 0.0111 & \it0.0015 & 0.0021 & \bf0.0012 & \bf0.0012 & 0.0024 \\
  \hline
  \hline
\end{tabular}
\end{table}

\begin{table}
\centering
\caption{\it $\eta$-MISE for all models under the sigmoidal and peak scenarios averaged over all 200 datasets. Bold-faced entries denote the lowest value at each Setting and Covariance Structure, Cov. Str., combination. Italicized entries denote the second smallest value. The abbreviations Ind., Exp. and, C. S. refer to Independence, Exponential, and Compound Symmetric, respectively.}
\label{t:eta}
\begin{tabular}{llccccccc}
  \hline
  \hline
 & \it Cov. & \multicolumn{2}{c}{\it B-Spline} & \multicolumn{2}{c}{\it O-Spline} & \multicolumn{2}{c}{\it Symmlets} & \it CLMM \\
\it Setting & \it Str. & $K = 5$ & $K = 10$ & $O = 2$ & $O = 4$ & $J = 6$ & $J = 8$ & \\ 
  \hline
  \hline
Sig. & Ind. & 0.0015 & 0.0010 & \bf0.0007 &\it0.0008 & \bf 0.0007 & 0.0023 & 0.1292 \\ 
 & Exp. & 0.0017 & 0.0012 & \it 0.0009 & 0.0010 & \bf0.0008 & 0.0025 & 0.1310\\ 
 & C. S. & 0.0016 & 0.0011 & \bf 0.0007 & 0.0009 & \it0.0008 & 0.0024 & 0.1292\\ 
   \hline
Peak & Ind. & 0.0077 & \it0.0010 & 0.0015 & \bf0.0008 & \bf0.0008 & 0.0017 & 0.1028 \\ 
  & Exp. & 0.0078 & 0.0011 & 0.0015 & \it0.0009 & \bf0.0008 & 0.0017 & 0.1035 \\
  & C. S. & 0.0078 & \it0.0011 & 0.0015 & \bf0.0008 & \bf0.0008 & 0.0017 & 0.1028 \\
  \hline
  \hline
\end{tabular}
\end{table}

In Table~\ref{t:beta}, the smallest average $\beta$-\emph{MISE} is bolded and second smallest is italicized. All basis functions perform reasonably well and similar to each other. However, the O-spline model with $O = 2$ knots has the smallest or second smallest $\beta$-\emph{MISE} under the sigmoidal setting while the O-spline model with $O = 4$ knots is smallest under the peak setting. When using $J = 6$ levels, the wavelet-based model also performs well for both settings while the B-spline model with $K = 10$ knots has the second smallest $\beta$-\emph{MISE} for the peak setting. For a given basis function, we see very little difference in $\beta$-\emph{MISE} across covariance structures. In Table~\ref{t:eta}, the CLMM produces $\eta$-\emph{MISE} values that are several orders of magnitude larger under both the sigmoidal and peak settings, regardless of the underlying covariance structure. Tabular $\beta$- and $\eta$-\emph{MISE} results from scenarios not presented here, as well as graphical depictions of estimated curves, are in the Supplementary Material. These additional results follow similar patterns to those we describe here.

While the underlying covariance structure does not have a large impact on MISE, it does affect coverage.  Table~\ref{t:cover} displays the joint coverage probabilities (JCP) for each OPFOR under the sigmoidal and peak settings. We calculate JCP by checking the number of times, out of 200, that the joint credible interval contains the entire true curve. Overall, O-spline models with $O = 4$ knots tend to produce JCPs that are the closest to the nominal coverage of 95\% while $B$-spline models with $K = 10$ knots tie it or are second closest. When using wavelets with $J = 6$ levels, the intervals exceed the nominal rate. But when using $J = 8$ levels, the intervals undercover. The exception is the decay setting where the wavelet-based models have the closest to nominal coverage and the spline-based models significantly underperform (see Supplementary Material). For all basis functions, JCP tends to be lower for more complex covariance structures. Tables of JCP's for scenarios not presented here and tables of point-wise coverage for all scenarios are in the Supplementary Material.

\begin{table}
\centering
\caption{\it Joint Coverage Probabilities or JCPs for the 95\% joint credible intervals for each OPFOR basis function under the sigmoidal and peak scenarios. Table values represent the percent of simulated datasets with intervals that completely contain the true curves. Bold-faced entries denote the closest value to the nominal level in each row. Italicized entries denote the second closest value. The abbreviations Ind., Exp., and C. S. refer to Independence, Exponential, and Compound Symmetric, respectively.}
\label{t:cover}
\begin{tabular}{lllccccc}
  \hline
  \hline
 &   & \multicolumn{2}{c}{\it B-Spline} & \multicolumn{2}{c}{\it O-Spline} & \multicolumn{2}{c}{\it Symmlets} \\
\it Setting & \it Cov. Str. & $K = 5$ & $K = 10$ & $O = 2$ & $O = 4$ & $J = 6$ & $J = 8$ \\ 
 \hline
 \hline
Sigmoidal & Ind. & 0.060 & \bf 0.955 & \it 0.915 & \bf 0.955 & 1.000 & 0.685 \\ 
 & Exp. & 0.015 & \it 0.905 & 0.835 & \bf 0.915 & \it 0.995 & 0.730 \\ 
 & C. S. & 0.055 & \it 0.925 & 0.875 & \bf 0.945 & 0.990 & 0.705 \\ 
 \hline
Peak & Ind. & 0.000 & \bf 0.935 & 0.105 & \it 0.925 & 1.000 & 0.840 \\
 & Exp. & 0.000 & \bf 0.915 & 0.105 & \bf 0.915 & \it 1.000 & 0.820 \\
 & C. S. & 0.000 & \it 0.920 & 0.135 & \bf 0.925 & 1.000 & 0.845 \\
 \hline
 \hline
\end{tabular}
\end{table}

\section{Analysis of Computer-Use Data}
\label{s:app}

We restrict our investigation to the first full monkey-year available, which consists of four seasons each roughly three months in length: the breeding season, winter, birthing season, and summer. By convention, the monkey-year begins in September with start of the breeding season and ends in August before the start of the next breeding season. For the 72 monkeys available for analysis, the amount of time spent using the touch-screens is binned on each day into the levels no, low, moderate, and high, resulting in 365 measurements per function for each monkey. No use corresponds to zero ms of computer use per day, low to between 559 ms and 50757 ms, medium to between 50771 ms 184206 ms, and high to between 184244 ms and 8207299 ms (for reference, tasks take between 1500 ms and 45000 ms to complete). The covariates of interest are time-invariant scalars: an indicator for whether the monkey is male, age (in months) at which the monkey received the RFID chip, and the monkey's dominance rank in the social group. Rank can be treated as either a numerical or categorical predictor, as raw ranks are often binned into low, medium, and high ranking categories. Summary statistics for each variable are in the Supplementary Material. Prior to modeling, we normalize age and the numerical rank to the 0 - 1 scale and center by the normalized mean.

The O-spline model performs well in simulation, particularly using $O = 4$ knots. Thus, we select this basis function for our analysis and build all possible main effect models. We then investigate the linearity of the numerical predictors and possible effect modifications. Table~\ref{t:crit} contains lppd and WAIC values for every model (see the Supplementary Material for DIC values). Focusing on the main effect models with a single covariate, the model with age alone has the lowest WAIC while the model with just the effect of being male has the second lowest suggesting that these are the two most important predictors. Notably, numerical rank has a lower WAIC than binned rank thus for models with more than one covariate, we use numerical rank. We evaluate quadratic effects in the two main effects models with the lowest WAICs. Across all of the main and quadratic effects models, the three best models are male $+$ age $+$ rank $+$ rank$^2$ (WAIC of 40404), male $+$ age $+$ rank (WAIC of 40554), and male $+$ age (WAIC of 41316). We investigate possible effect modifications in these three models.

\begin{table}
	\centering
	\caption{\it Log posterior predictive density (lppd) and Watanabe-Akaike Information Criterion (WAIC) for all models under consideration. Models with values of ``$-$'' had unstable estimates. Italicized values are the lowest WAICs within Model Type while the bolded value is the lowest WAIC overall.}
	\label{t:crit}
	\begin{tabular}{llcc}
		\hline
		\hline
		\it Model Type & \it Covariates & \it lppd &\it WAIC \\
		\hline
		\hline
		Main Effects & male & -17080 & 45903\\
		  & age & -16489 & \it 43718\\
		  & rank & -17200 & 47344\\
		  & binned rank & -17187 & 47559\\
		  \cline{2-4}
		  & male, age & -15853 & \it 41316\\
		  & male, rank & -17192 & 45278\\ %
		  & age, rank & -16531 & 43473\\
		  \cline{2-4}
		  & male, age, rank & -15724 & \it 40554\\
		 \hline
		Quadratic Effects & male, age, age$^2$ & $-$ & $-$\\
		& male, age, rank, rank$^2$ & -15604 & \it 40404\\
		 & male, age, rank, age$^2$ & -15626 & 40853\\
		 \hline
		 Effect Modifications & male, age, male $\times$ age & -15970 & \textbf{\emph{38063}}\\ 
		  & male, age, rank, male $\times$ age & -15761 & 39709 \\
		  & male, age, rank, male $\times$ rank & -15781 & 40267 \\
		 & male, age, rank, rank$^2$, male $\times$ age & -15651 & 39466\\
		  & male, age, rank, rank$^2$, & \multirow{2}{*}{$-$} & \multirow{2}{*}{$-$} \\
		  & \hspace{25pt}male $\times$ rank, male $\times$ rank$^2$ \\
		\hline
		\hline
	\end{tabular}
\end{table}

Including the interaction of male and age improves the fit of all three models. The interaction of male and rank does not improve model fit or results in unstable estimates. From the effect modification models, we find the model with male, age, and the interaction of male and age to have the lowest WAIC, 38063, making it the best fitting model overall. We consider this our final model. A sensitivity analysis to the choice of basis function using B-splines with $K = 10$ knots suggests the same final model when using WAIC as our selection criterion (see Supplementary Material).

\begin{figure}
\centering
\includegraphics[height = 1.75in, width = 2.25in]{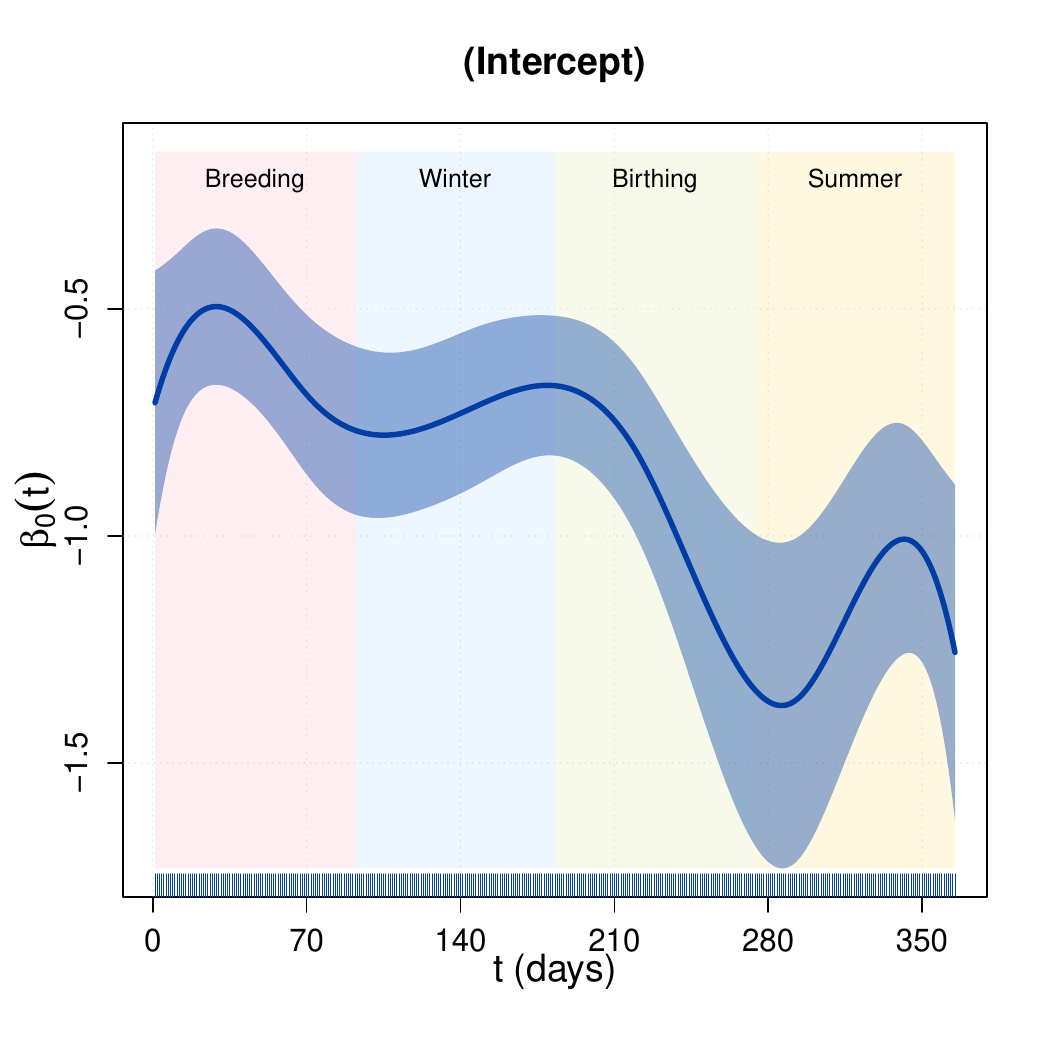}
\includegraphics[height = 1.75in, width = 2.25in]{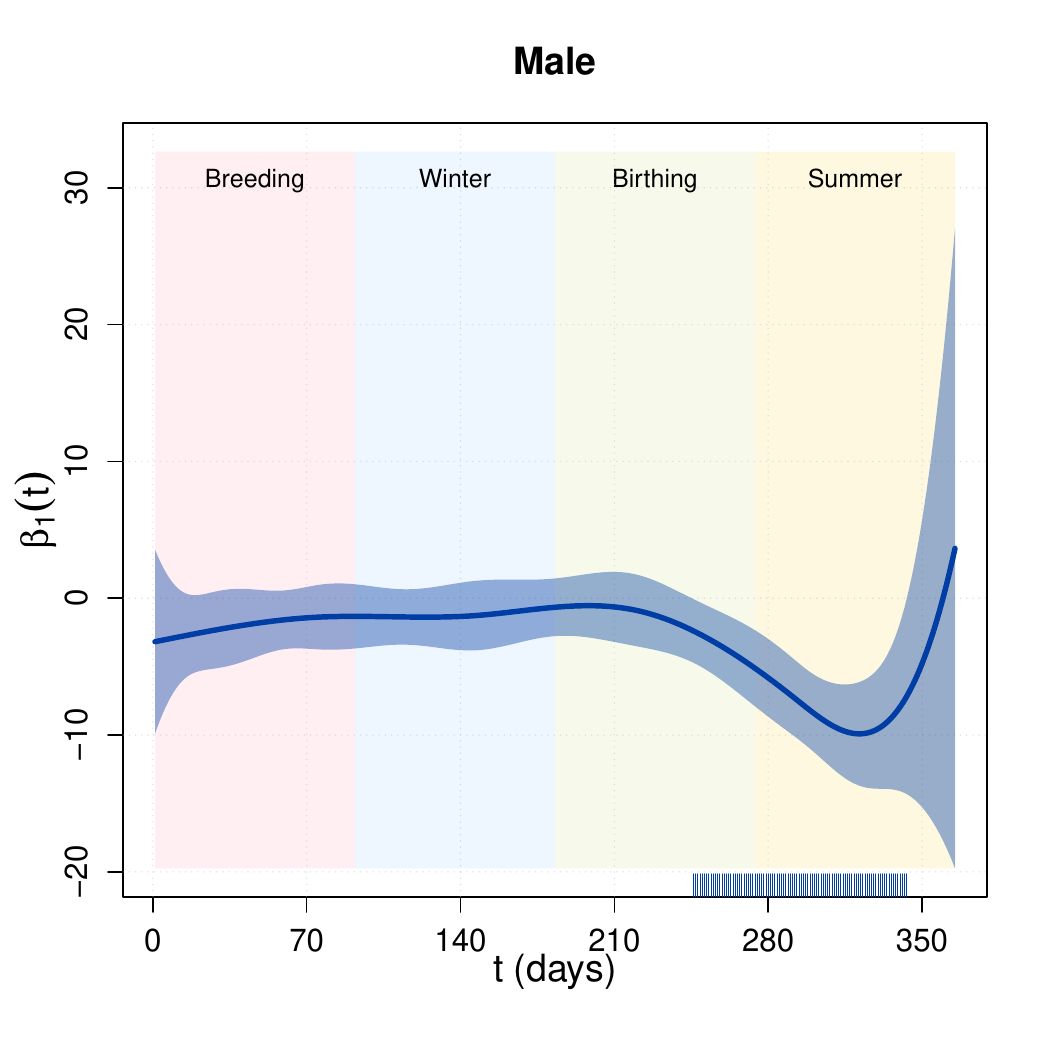}
\includegraphics[height = 1.75in, width = 2.25in]{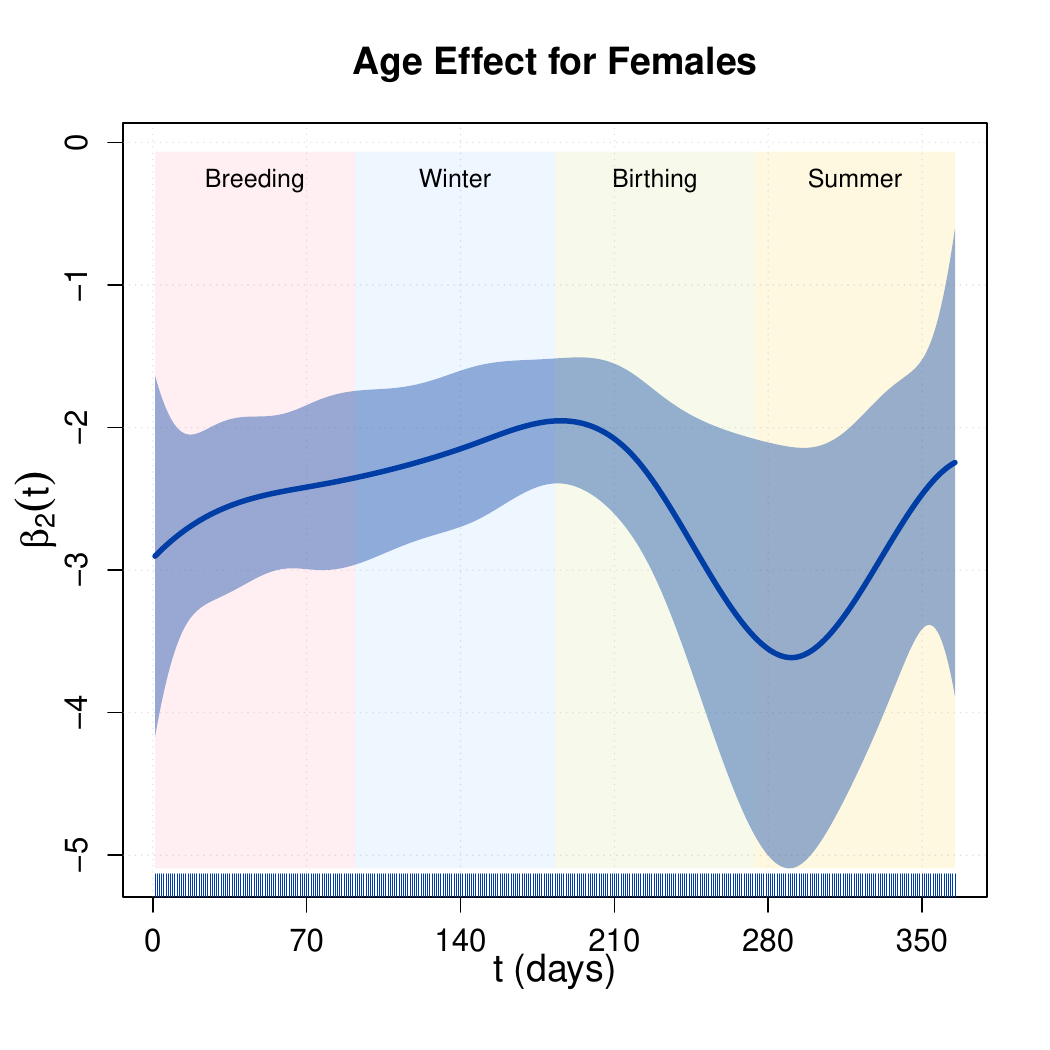}
\includegraphics[height = 1.75in, width = 2.25in]{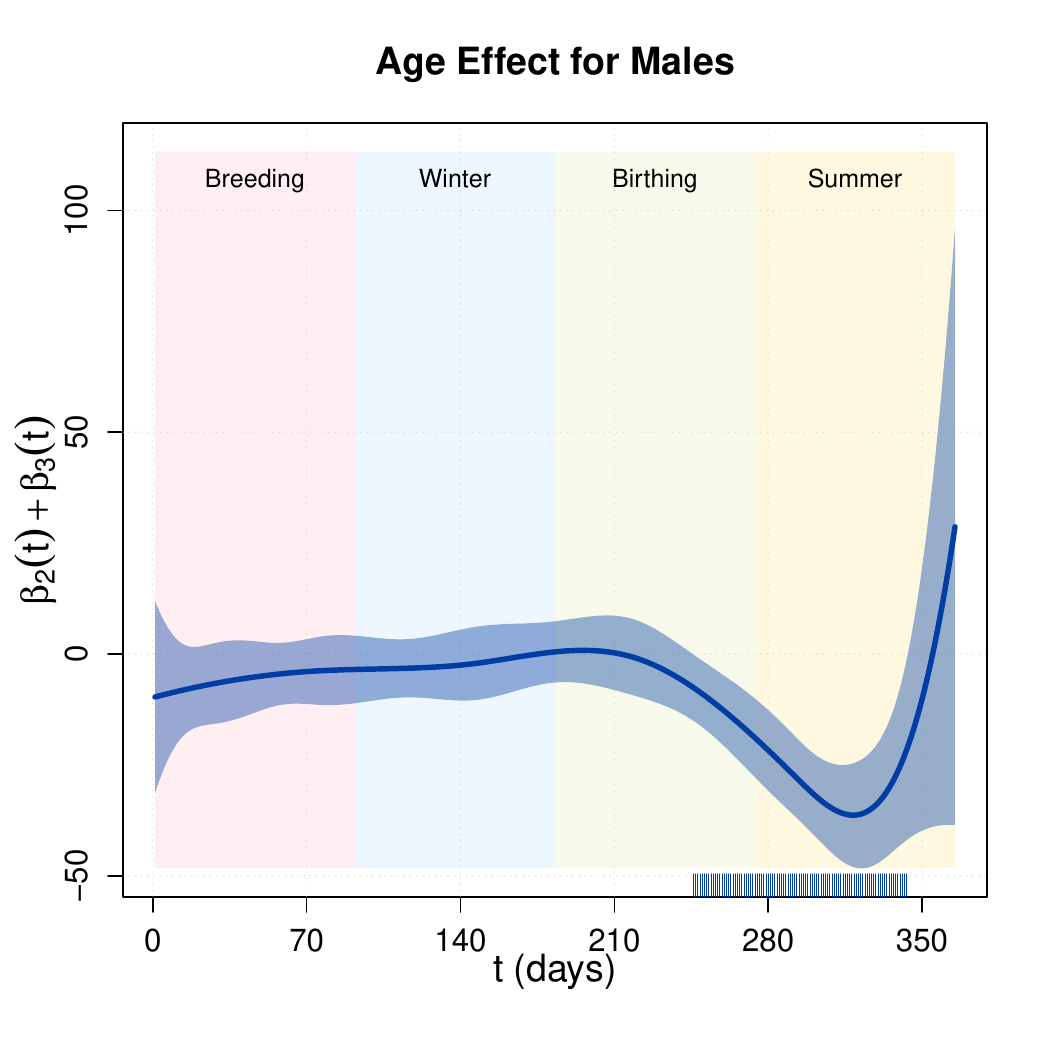}
\caption{\it Posterior estimates and joint intervals of demographic factor effects on computer-use, model fit using O-spline with $O = 4$ knots. The solid blue curves indicate the posterior mean of $\beta(t)$ while the blue bands denote the 95\% joint credible interval. The rug in each graph denotes coefficients deemed significant by the joint intervals. The seasons are color-coded and labeled. This figure appears in color in the electronic version of this article.}
\label{f:osp4}
\end{figure}

The final model is, in the augmented space, $Y^*_i(t) = \beta_0(t) +  \text{male}_i \beta_1(t) + \text{age}_i \beta_2(t) + \text{male}_i\times\text{age}_i\beta_3(t) + E_i(t)$. All parameter estimates were judged to have converged on 15,000 retained posterior samples after a warm-up of 15,000---the additional covariates and, therefore, parameters require longer chains to converge than did the simulated data sets, see Supplementary Material for diagnostics. Estimated curves (solid blue lines), joint credible intervals (blue bands), and non-zero regions of each curve by interval (blue rug) are in Figure~\ref{f:osp4}. For interpretability, each figure presents the effect of changing the covariate after adjusting for all other covariates in the model. Further, we separately display the effect of age for males and for females using the interaction.

Results in Figure~\ref{f:osp4} indicate seasonal variation in the effects of age and sex on touch-screen computer use. After adjusting for age, male computer use is less than females from the end of the birthing period through the summer, but there is no sex difference in activity during the breeding season, winter, and or most of the birthing season. Age is a more important predictor of computer use for female monkeys than for males. The effect of age for females is consistently less than zero suggesting that older females use the computers less than younger females, regardless of the season. For males, age was only related to computer-use beginning in the latter part of the birthing season and through the summer, with older males using the computer less than younger male monkeys during this time. By construction, the cut-points were fixed at the same levels across all days, thus we fixed the values of the cut-points in the model. A sensitivity analysis where we sample the cut-points produces similar estimates (see Supplementary Material).

In the longitudinal regression analysis where, instead of treating each monkey's vector of ordinal variates as a functional outcome, we assume the vectors contain repeatedly sampled longitudinal outcomes and fit a cumulative-link mixed-effects model (CLMM) with a probit link. To account for repeated sampling, we include a random subject-specific intercept as well as a random subject-specific slope on time. We obtain parameter estimates using the \texttt{clmm} function from the \texttt{ordinal} package in \texttt{R} \citep{Ordinal2019}. Results from the CLMM are presented in the Supplementary Material. Like the OPFOR, this approach indicates a significant effect of age, modified by sex, on computer-use. However, the CLMM does not provide the same clarity regarding the seasonality of these effects that the OPFOR produces. For example, the CLMM indicates that males interact with the computer systems less than females. However the OPFOR reveals that this difference is only present during the birthing and summer seasons, indicating that the sex difference is a seasonal effect rather than a general tendency for males to interact with the systems less than females.

\section{Discussion}
\label{s:disc}

Although recent advances in functional outcome regression consider non-Gaussian outcomes, more work is needed to expand the available methodology. For ordinal functional outcomes, the literature is quite sparse consisting of only one limited investigation and no publicly available code. Given this and to better analyze the computer-use data, we introduce a fully Bayesian functional outcome regression for functional $L$-level ordinal variates which we refer to as OPFOR. The primary novelty of our approach lies in the use of the data-augmented space to represent the ordinal functional outcome as a latent Gaussian functional outcome. This representation provides a flexible modeling framework using a variety of basis functions including B-splines and discrete wavelets, which are commonly used in the literature, as well as O-splines.To our knowledge, O-splines have not previously been implemented in generalized function-on-scalar regression. Building on the frameworks of \cite{MorrisCarroll2006} and \cite{Goldsmith2016}, both spline- and wavelet-based formulations of the model account for within curve correlations in the basis-transformed latent space either via the use of fPC or the whitening property of the DWT and therefore account for within curve correlations in the ordinal variates. OPFOR can also handle a potentially large number of subjects and any combination of numerical and categorical scalar covariates, as demonstrated by our analysis of the computer-use data. Finally, we formulate several Bayesian model selection criteria for use in determining the model of best fit.

Our simulation study investigates the operating characteristics of OPFOR and compares it to a standard ordinal longitudinal model for a reasonable sample size of $N = 40$ under four ``true'' curves and three underlying covariance structures. While the O-spline models perform best in terms of $\beta$-\emph{MISE}, we show that the model performs well in estimation under any choice of basis function. Varying the underlying covariance structure from independent to time-invariant to time-varying does not have a significant impact on estimation which illustrates the ability of both the DWT and fPC to capture within curve correlation in the latent space. Since the covariance structure does impact the JCP, we prefer the use of O-splines with $O = 4$ knots. B-splines using $K = 10$ knots are a plausible second choice given their consistently near-nominal coverage levels. By comparison, the CLMM performs poorly under all simulated scenarios and is computationally less efficient than the OPFOR when using O-splines or B-splines.

The analysis of the computer-use data using OPFOR reveals important trends that the standard longitudinal analysis misses. For example, a previous longitudinal regression analysis indicates that females engage with the touch screen systems significantly more than males \citep{Gazes2019}, suggesting that females might be better research subjects in this testing environment than males. However, the present analyses clarify this effect, indicating that the sex effect is only present in the summer months, not during the rest of the year. Similarly, previous analyses indicate no effect of age on touch-screen use \citep{Gazes2019}, but the results of the present analyses indicate that the effect of age on performance is more nuanced. In males, this lack of an age effect is present for most of the year, but in the summer, younger monkeys engage with the systems more than older monkeys. These seasonal sex and age effects suggest that something is happening in the lives of adult male monkeys in the summer months that interferes with their ability to or interest in participating in cognitive testing. What this behavioral change may be remains an empirical question. Practically, future studies of sex differences in cognition in this population should focus data collection during the fall, winter, and spring to obtain comparable touch screen engagement from male and female subjects. 

%
%
%
%

\end{document}